*Type of the Paper (Article, Review, Communication, etc.)*

# Adaptive Cybersecurity: Dynamically Retrainable Firewalls for Real-Time Network Protection

Sina Ahmadi[1]

[1] The National Coalition of Independent Scholars (NCIS), sina0@acm.org

**Abstract:** The growing complexity of cyberattacks has necessitated the evolution of firewall technologies from static models to adaptive, machine learning-driven systems. This research introduces "Dynamically Retrainable Firewalls", which respond to emerging threats in real-time. Unlike traditional firewalls that rely on static rules to inspect traffic, these advanced systems leverage machine learning algorithms to analyze network traffic patterns dynamically and identify threats. The study explores architectures such as microservices and distributed systems for real-time adaptability, data sources for model retraining, and dynamic threat identification through reinforcement and continual learning. It also discusses strategies to improve performance, reduce latency, optimize resource utilization, and address integration issues with present-day concepts such as Zero Trust and mixed environments. By critically assessing the literature, analyzing case studies, and elucidating areas of future research, this work suggests dynamically retrainable firewalls as a more robust form of network security. Additionally, it considers emerging trends such as advancements in AI and quantum computing, ethical issues, and other regulatory questions surrounding future AI systems. These findings provide valuable information on the future state of adaptive cyber security, focusing on the need for proactive and adaptive measures that counter cyber threats that continue to evolve.

## 1. Introduction

The rapidly evolving cybersecurity landscape demands innovative solutions that surpass the limitations of traditional defense systems' capabilities. Conventional firewalls often fail to protect against sophisticated threats such as zero-day attacks, polymorphic malware, and advanced persistent threats (APTs) [1]. Emerging shapes of organizational IT architecture, such as hybrid clouds, IoT devices, environments, and segmented networks, further exacerbate these challenges. In this regard, firewalls capable of retraining on the fly represent a revolutionary development made possible by machine-learning-based threat detection systems. This paper covers the design, deployment, and utilization of these firewalls, with a focus on addressing scalability, performance, and integration in modern complex networks.

*1.1. Background and Context*

The limitations of static firewalls stem from their reliance on predefined rule sets, which cannot anticipate or respond to the rapid evolution of cyber threats [2]. Modern attacks are more complex and require organizations to use a system that can react to real-time changes. This is served by dynamically retrainable firewalls, which employ machine-learning models that analyze network data and adapt firewall settings to emerging threats. This approach enhances the level of recognizing deviations and accelerates the detection execution time, minimizing the destruction of breaches. The emergence of micro-segmentation and Zero Trust architectures has further emphasized the need for dynamic adaptability. These architectures continuously monitor and validate network behavior, requiring firewalls that can retrain and adjust in real time to maintain robust security.

*1.2. Significance of Dynamically Retrainable Firewalls*

Dynamically retrainable firewalls are transforming the field of cybersecurity [3]. Unlike static firewalls, which rely on fixed configurations, these systems continuously adapt by incorporating fresh data to identify and respond to threats. This capability is essential for detecting unknown threats that often escape standard security measures. By leveraging reinforcement learning and continual learning, dynamically retrainable firewalls can update their parameters to address modern threats. It not only improves the ability to identify hazards but also helps to reduce the extent to which threat identification and monitoring depend on manual repetitions, which are slow and susceptible to mistakes.

Scalability is another critical advantage of dynamically retrainable firewalls [4]. As the organization or its network's size increases, the volume of data requiring analysis grows rapidly. Conventional firewalls often struggle under such expectations, causing network throughput stagnation and escalation risks. Dynamically reconfigurable firewalls, in contrast, are envisioned to run in high-traffic zones where they must offer optimal performance, often through deploying distributed architectures and efficient algorithms [5]. This kind of scalability makes them practical for cloud and hybrid networks since the access demand can be highly variable.

Integrating these firewalls with modern network architectures further enhances their utility. In Zero Trust, the tenant is "never trust, always verify," hence constant monitoring and dynamic threat detection are pursued [6]. Firewalls can be dynamically retrained to ratchet up with this approach and offer the real-time change needed to implement the Zero Trust model properly. Furthermore, compatibility with infrastructures like hybrid and multi-cloud is crucial to seeing that various segments of an organization's network retain a similar security position.

*1.3. Objectives and Research Questions*

This research aims to advance the understanding and implementation of dynamically retrainable firewalls by addressing several key questions.
1. How can scalable architectures support real-time adaptation without introducing latency?
2. Which machine learning techniques are most effective for continuous threat identification and categorization?
3. How do these firewalls fit into the networks' current networks, including hybrid and multi-cloud environments?
4. What metrics should be used to evaluate their performance in production settings?

To answer these questions, this research aims to frame the ways of applying dynamically retrainable firewalls and provide organizations with the necessary tools to improve their levels of cyber security.

*1.4. Scope of the Research*

The scope of this research encompasses the key components of dynamically retrainable firewalls, including their design, implementation, and evaluation. The Design and Architecture section will design the mechanisms based on scalable XCAs containing the ML models that may be retrained in real time. The Dynamic Threat Detection section discusses advancements in machine learning, namely reinforcement and continual learning, to identify unknown and constantly evolving threats. The Performance Optimization section focuses on maintaining real-time performance with minimal latency and resources consumption. The Integration with Modern Networks sub-topic examines how these firewalls can be integrated into Zero Trust architectures, cloud, and hybrid environments. Finally, the Evaluation and Applications section evaluates the feasibility and impact of dynamically retrainable firewalls through experiments.

## 2. Literature Review and Background

*2.1. Foundational Concepts*

2.1.1. Overview of Firewalls: Static vs. Adaptive

Firewalls have traditionally served as the first line of defense for networks, connecting safe internal networks with hostile environments [7]. Historically, stateful firewalls operate based on a set of administrator-defined rules, permitting traffic only as specified by these guidelines. These firewalls remain relevant in addressing known threats with simple, predictable traffic patterns but are inadequate against the complexity of modern cyber threats. Explicit firewalls depend on manual updating; hence, they are susceptible to zero-day, polymorphic, and adaptive cyber threats. In contrast, adaptive firewalls are far from traditional firewalls. These systems utilize adaptive algorithms to dissect flow and interaction occurring on the network at any given time and determine deviations, frequent uses, and security risks. Unlike their static counterparts, adaptive firewalls require minimal human intervention, transitioning from a reactive to a proactive security model [8]. They are particularly well-suited for segmented networks, hybrid cloud infrastructures, and environments where IoT devices play a central role. Table 1 shows static firewalls vs dynamic firewall features.

| Feature | Static Firewalls | Dynamically Retrainable Firewalls |
|---|---|---|
| Adaptability | Fixed rules, and manual updates required | Learns from real-time data, auto-adapts |
| Threat Detection | Limited to known threats | Identifies unknown threats using ML |
| Scalability | Struggles in high-traffic environments | Scalable with distributed architectures |
| Performance Optimization | Minimal computational demand | Requires real-time data and processing |
| Integration | Limited adaptability to modern networks | Seamlessly integrates with Zero Trust, hybrid, and cloud environments |

**Table 1.** Comparison of Static and Dynamic Firewalls.

2.1.2. Role of Machine Learning in Cybersecurity

Machine learning has become one of the most influential paradigms in cybersecurity due to its ability to analyze numerous data points, reveal patterns, and even accurately forecast threats [9]. Conventional rule-based solutions require programmatic development and cannot handle new, unknown threats. On the other hand, ML-based systems will be able to learn from the data and defend themselves from threats in real time.

Supervised learning algorithms enable systems to be trained on reliable databases of identified threats to enhance the techniques of identifying anti-social actions [10]. Unlike unsupervised learning, this helps identify unknown characteristics or outliers in network traffic that may relate to new attacks. Reinforcement learning adds to these systems by continuously adapting them to provide better performance depending on feedback in constantly evolving networks.

In the case of dynamically retrainable firewalls, ML offers the computation support required for constant detection and adaptable response. They achieve their robust security statuses by analyzing the traffic data, recognizing the patterns, and retuning the models on new threats discovered. The inclusion of ML into adaptive firewalls means they can deal with modern threats, providing the scale and flexibility needed due to the multiple firewalls used in various networks [11]. Figure 1 represents the core of the ML system, learning patterns and identifying anomalies in the network traffic.

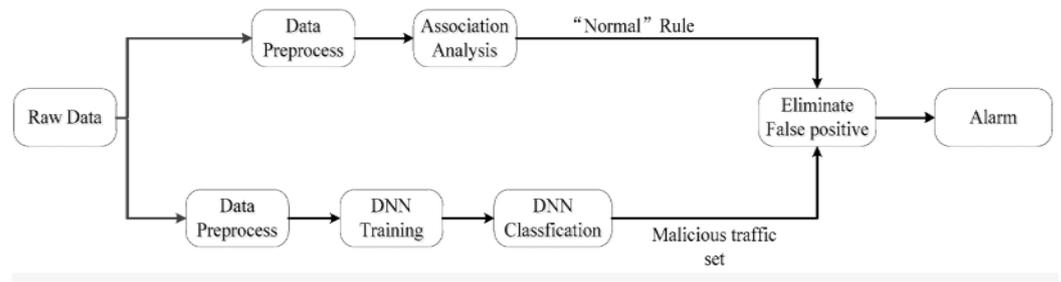

**Figure 1.** ML in Cybersecurity.

*2.2. Existing Research*

2.2.1. Review of Past Work on Dynamic Threat Detection and Adaptive Models

Dynamic threat detection has garnered significant interest, with researchers focusing on developing solutions capable of detecting and countering threats in real time using the power of ML and AI [12]. Early research highlighted the potential of anomaly detection algorithms, which compare current network behavior against baseline data to identify unusual or suspicious activity. Though helpful, these methods were not always very accurate, mainly because they would sometimes yield high rates of false positives and were not easily expansible.

Recent developments have produced more complex methods, including ensemble and deep learning. Ensemble learning combines multiple models to achieve higher detection efficiency, while deep learning employs neural networks to process and analyze large datasets [13]. These techniques have been implemented earlier in areas, such as intrusion detection systems and malware classification.

Dynamic firewalls, in particular, have benefited significantly from these advancements [14]. Case studies in enterprise environments demonstrate the advantages of adaptive firewalls, indicating improvements in response time and threat detection accuracy. For instance, one investigated the application of reinforcement learning for determining firewalls' optimal configuration that adapts to the network's feedback. This gave visible results in the sense of reduced damage from cyberattacks and better preparedness for new and more complex threats. Figure 2 depicts the ever-evolving landscape of cyber threats, emphasizing the dynamic nature of modern security challenges.

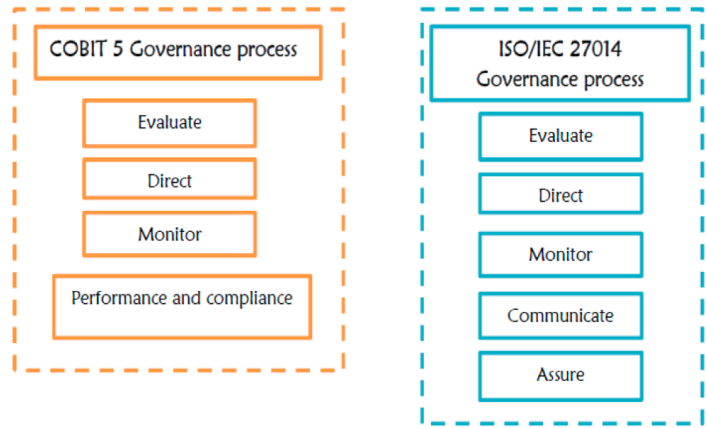

**Figure 2.** Dynamic Threat Detection and Adaptive Models.

2.2.2. Case Studies Highlighting Successes and Limitations

Numerous case studies illustrate the successes and limitations of dynamically retrainable firewalls. For instance, a large financial firm implemented an adaptive firewall system supported by machine learning algorithms. The firewall demonstrated impressive performance, particularly in addressing phishing threats, reducing their occurrence to below 10% [15]. Similarly, cloud service providers have also embraced adaptive firewalls for their infrastructures, ideally showcasing their capacity and relevance in high elasticity and capacity demands.

Despite these successes, challenges persist. A major issue is the computational load required for real-time model retraining, which complicates network operations and reduces overall speed. Moreover, adaptive firewalls typically use big data as their learning basis while raising such issues as data privacy and model prejudice [16]. These limitations underscore the need for ongoing innovation to address the essential compromises at the heart of any implementation of the three Ps of performance, scalability, and flexibility.

*2.3. Gaps in Current Knowledge*

2.3.1. Challenges with Scalability

Scalability remains a primary challenge for dynamically retrainable firewalls. Modern network conditions manage tremendous traffic, especially in business and cloud domains. The real challenge is ensuring adaptive firewalls can manage such an amount of traffic without degrading the system. Current solutions often fall short as the number of endpoints escalates. In the case of integrated and multiple cloud networks, traffic tends to shift frequently. Efforts to address scalability have focused on distributed architectures, which distribute computational workloads across multiple nodes [17]. Although appealing, these approaches create challenges, including maintaining data integrity and coherency between nodes. In general, creating mechanisms that permit dynamically retrainable firewalls to operate on enormous amounts of data swiftly, without experiencing a decline in precision, is paramount.

2.3.2. Latency and Resource Efficiency

Latency is a significant issue in real-time environments, where even minor delays can pose severe security risks [18]. Retraining the machine learning (ML) models in real time demand substantial computational resources, which increases the detection load and impacts overall speed and accuracy. Low false negative rates indicate additional processes to ensure there are no omitted false negatives that accumulate latency. Such delays may allow threats to navigate around several layers of protection, weakening the existing network's defense. While high flow conditions are more challenging to handle, trade-offs are involved in generating immediate responses to queries while incurring a computational cost of adaptive retraining. Reducing latency is paramount as a dynamically reconfigurable firewall should be able to give accurate protection when in real-time and with as little delay as can be needed in handling today's complicated cyber threats [19].

Resource efficiency is another pressing concern, especially in resource-constrained environments like Internet of Things (IoT) networks, where hardware limitations present significant challenges. [20]. Subsequently, the quest to develop novel machine learning algorithms that can run on restrictive computing environments, such as embedded systems with minimal resources, remains crucial. Application development approaches like model compression are employed to prune the learning model to less computational requirements and perform in less time. In addition, edge computing emerges as a promising solution, which implies shifting part of a computational work closer to the data source, avoiding multiple uses of centralized computing [21]. These approaches allow adaptive firewalls to maintain high efficiency throughout restricted resource availability, and security solutions can be upgraded and executed concurrently with ordinary network functioning. Figure 3 illustrates network latency, depicting the 1.4 seconds it takes for a user request to travel from a PC to a data center and for the server's response to return.

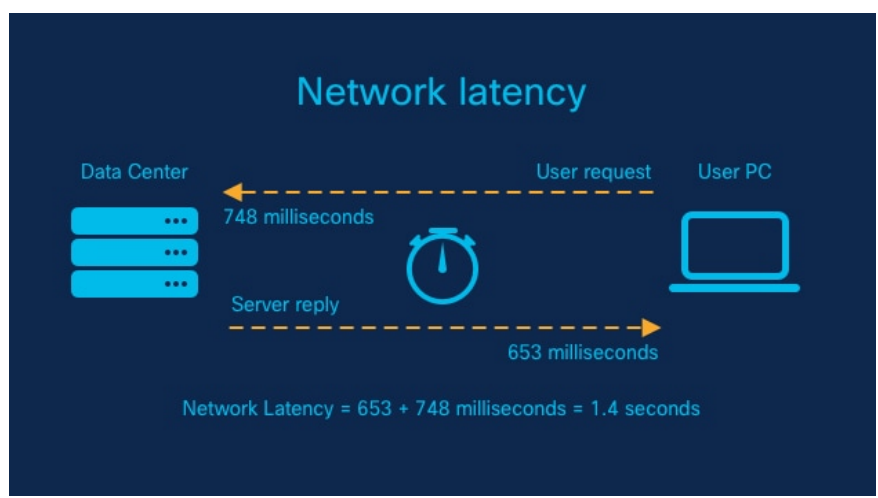

**Figure 3.** Latency Efficiency.

2.3.3. Threat Evolution

The rapidly evolving nature of cyber threats poses a significant challenge for dynamically retrainable firewalls. Cyber attackers continuously invent new methods to bypass traditional protection mechanisms, hence necessitating advanced systems. Another drawback of the ML algorithms is their strong performance against threats similar to those previously observed in the network system but limited generalization to novel threats [22]. Increasing the generalization capability of ML models is a highly relevant problem for the long-term effectiveness of the systems.

*2.4. Theoretical Frameworks*

2.4.1. Integration of Zero Trust Principles

The Zero Trust security model has gained widespread adoption as a framework for modern cybersecurity, emphasizing the principle of "never trust, always verify." Whereas other security models adhere to the idea that security starts at the perimeters of the network, Zero Trust considers the network to be compromised. He constantly requires proper credentials from every user and device [23]. Dynamically retrainable firewalls align well with Zero Trust principles, providing real-time monitoring and adaptive responses to potential threats. These systems increase the Zero Trust policy precision by actively monitoring network traffic and tweaking settings. For instance, an adaptive firewall can identify when an authenticated user behaves suspiciously, leading to other authentication steps or the user being barred from accessing some resources [24]. This dynamic approach means that even if an attacker secures the first foothold, their privileged move in the network is hugely restricted.

2.4.2. Segmented Network Architectures

Segmented network architectures further complement the role of dynamically retrainable firewalls [25]. By dividing networks into smaller, isolated segments, organizations can limit the damage caused by security breaches and improve threat detection. This segmentation prevents threats from spreading freely across the network. Firewalls installed at segment boundaries can enforce security measures proportional to each segment's needs. For example, different segments may be dedicated to patient records, medical devices, and administrative systems in a healthcare network. An adaptive firewall can establish robust access control parameters firmly for each segment to prevent several segments from being compromised [26]. This segmentation not only optimizes security but also rationalizes the introduction of compliance with legal standards and regulations, such as those related to data protection.

## 3. Problem Definition

The rising complexity of cyberattacks presents a significant challenge to traditional cybersecurity solutions [27]. As threats evolve at a breakneck pace, so do the tools used by the attackers. For example, advanced persistent threats, zero-day exploits, and polymorphic malware, among others, have proven particularly challenging for traditional security mechanisms to counter. Firewalls generally fall into two broad categories: static and semi-dynamic firewalls, which use rule-based systems and signature-based detection techniques, which offer limited capabilities for addressing emerging threats. While these firewalls provide protection against known attacks, they are vulnerable to novel tactics if they lack adaptive methods, leaving systems exposed to such evolving threats.

These restrictions on static and semi-dynamic firewalls raise a pressing demand for more advanced, dynamic, and preemptive solutions. Another good approach is dynamically retrainable firewalls using machine learning for firewall updates and real-time learning of new attack types. Yet, this is not without difficulty. Continuous reinforcement learning is a very computationally intensive process that adds delay to the detection and response time that the firewall uses to block threats. Hence, threats may sneak through the firewall.

This research seeks to answer several essential questions. First, how can the fact that firewalls are retrainable during network transactions enhance how the networks detect and counter new dangers? This question aims to establish whether current reinforcement learning and continual learning approaches can be employed to adapt firewalls to detect other forms of threats that were not initially programmed. Second, what is the cost regarding system capability for real-time retraining? While firewalls evolve according to the existing threats, the need to assess every transaction will mean that specific issues related to computational load may slow down the system, especially where the traffic is high. Hence, these must be addressed to create firewalls capable of incorporating brilliant resistances in real time without compromising performance.

## 4. Research Agenda

The research agenda focuses on advancing dynamically retrainable firewalls by exploring key areas such as scalable architectures, dynamic threat detection, and testing methodologies. A central focus area will be building generalized firewall formation with learning models allowing real-time retargeting. This involves making a design that can handle the growing traffic of your network while keeping the firewall flexible to future threats. Thus, by analyzing scalable architectures, the research will guarantee that the firewall can function effectively at various organizational scales and types of networks, ranging from small businesses to cloud networks.

Another critical research direction is dynamic threat detection, with an emphasis on reinforcement learning and continual learning. These approaches allow firewalls to adapt to real-time network traffic and detect novel or previously unseen attacks. The research will investigate strategies for successfully integrating these techniques into firewall systems to enhance their performance and accuracy.

This methodological approach includes a proposed framework for testing and evaluating dynamically retrainable firewalls. This framework will center on fundamental performance indicators like latency and accuracy to determine the firewall's ability to identify fresh threats and offer solutions without slowing down the system. By assessing the five metrics, the research will give a practical analysis of the benefits and shortcomings of using dynamically retrainable firewalls in various settings.

## 5. Discussion

*5.1. Design and Architecture: Scalable, Distributed Systems for Real-Time Adaptation*

The design and architecture of dynamically retrainable firewalls are fundamental to their ability to offer real-time, adaptive protection in modern networks. A distributed architecture with scalability enables the firewall to handle large volumes of traffic without straining the system's computational resources. This scalability is increasingly required due to the growing size and complexity of networks, which traditional, centralized firewalls can no longer support. A distributed system divides the load across several devices or nodes, enabling concurrent data processing, real-time analysis, and adjustments in high-traffic conditions.

The architecture must also support the continuous retraining of machine learning (ML) models. This requires incorporating data acquisition systems, model development systems, and inference processes that can update the firewall's detection and response capacities in line with current threat intelligence. Data sources for retraining comprise traffic logs, Intrusion Detection Systems (IDS), and feeds from threat intelligence, which provide the necessary input for identifying emerging patterns of observed malicious activities. For instance, network traffic may reveal a previously unknown communication pattern that the firewall must learn about as a new type of attack. Equation 1 represents traffic or computational load distribution across multiple nodes in a system, where $T$ is the total traffic and $L_i$ is the load handled by the *i-th* node.

$$C_t = \sum_{i=1}^{n} \frac{T}{L_i} \qquad (1)$$

Furthermore, real-time adaptation of certain subsystems within the firewall must ensure quick and easy interaction between them [28]. The system should be designed so that the continuous update of the ML models does not interfere with the firewall's core services. Managing the tension between scalability and flexibility involves complex architectural solutions such as microservices, containers, and cloud-based solutions, which can be rapidly adjusted based on the requirements of the network.

*5.2. Data Sources and Integration Methods for Model Retraining*

Effective dynamic retraining of firewalls relies heavily on the quality and diversity of data sources. Network traffic is a critical data source used to train models for attack identification [29]. This is why it is beneficial to gather data from multiple layers of the network stack, which helps in understanding traffic flow better and, thus, in predicting potential threats more accurately. This information can originate from many network points, including routers, switches, and firewalls, among others, and end devices.

Another essential source of data is threat intelligence feeds. These feeds ensure that information on known threats, potential attack methods, and IOCs is fed into the system in real time [30]. Due to retraining with threat intelligence, the firewall will constantly be updated with the latest attack methods and risks. Integrating threat intelligence allows the firewall to adapt quickly to emerging threats, significantly enhancing its ability to respond to zero-day attacks and other unknown threats.

Integrating data from these sources requires sophisticated preprocessing, normalizing, and combining diverse datasets. For instance, feature engineering and data augmentation are helpful when preparing the data for the ML models [31]. In addition, the integration needs to be explicitly designed to anticipate data latency and allow retraining activities to happen without impacting the firewall's operation. This encompasses the development of fruitful lines of continuous data feed and processing, along with timely model updates that can add robust dynamism to the system flow while maintaining optimal performance.

*5.3. Dynamic Threat Detection: Algorithms and Approaches for Handling Unknown Threats*

One of the key advantages of dynamically retrainable firewalls is their ability to detect and mitigate previously unknown threats—an area where traditional firewalls fall short. The integration of anomaly threat detection is critical, especially with the advanced types of threats used by modern malware, like polymorphic viruses, zero-day exploits, and APTs that signature-based detection cannot quickly identify [32].

Machine learning algorithms, particularly supervised and unsupervised learning models, are at the heart of dynamic threat detection [33]. In supervised learning, algorithms are trained to minimize the differences between new features in network traffic and existing datasets of standard and malicious traffic. However, these models remain by the training data used and need constant updates to remain efficient. On the other hand, unsupervised learning can detect abnormal network traffic behavior without labeled data. These models identify new, previously unknown threats by analyzing the variance of the regular network traffic. Equation 2 meausres precision, which is the proportion of true positives detected (*TP*) vs the summation of true positives and false positives (*FP*).

$$P_a = \frac{TP}{TP + FP} \qquad (2)$$

Another promising approach for dynamic threat detection is reinforcement learning (RL). In RL, the system receives feedback from its environment and adapts by learning from errors. This technique enables firewalls to update their response mechanisms as new threat permutations emerge. Reinforcement learning is particularly beneficial in environments with high variability, where traditional threat detection methods may struggle to identify new and emerging attack strategies. Figure 4 illustrates how threat intelligence feeds work by gathering data from sources like open-source intelligence, honeypots, malware analysis, and threat actor tracking.

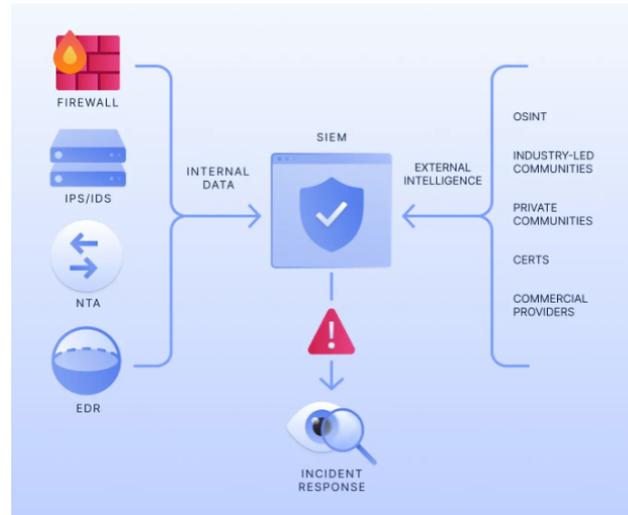

**Figure 4.** Threat Intelligence Feeds into Emerging Cyber Threats.

Case studies on dynamic threat detection have significantly improved detection rates, especially in identifying zero-day attacks. For instance, when reinforcement learning was applied in intrusion detection, the system successfully identified new attack patterns that previous methods could not detect. These improvements are significant for defending against highly advanced and constantly evolving cyber threats, highlighting the increasing need for robust firewall protection today. Table 2 compares the strengths and weaknesses of different machine learning algorithms.

| Technique | Strengths | Weaknesses | Use Case |
| --- | --- | --- | --- |
| Supervised Learning | High accuracy for known patterns | Requires labeled datasets | Known malware detection |
| Unsupervised Learning | Detects unknown patterns | Higher false positives | Anomaly-based intrusion detection |
| Reinforcement Learning | Learns from feedback, dynamic adaptability | Computationally intensive | Real-time adaptive firewall adjustments |

**Table 2.** Comparison of machine learning algorithms.

*5.4. Case Studies Demonstrating Detection Improvements*

Several real-world case studies have demonstrated the effectiveness of dynamically retrainable firewalls in improving threat detection. For instance, the use of adaptive firewalls in large-scale cloud environments successfully identified new, previously unknown attacks [34]. The system updated machine learning algorithms with flow analysis and threat intelligence feeds, enabling the firewall to identify threats that conventional, signature-based means could not identify.

A similar case occurred in a financial institution that deployed a machine learning firewall integrating supervised and unsupervised models for behavior recognition [35]. The firewall detected an advanced persistent threat (APT) that had eluded other detection tools for weeks. The firewall's ability to repeatedly train the models allowed it to uncover new attack strategies that had not been identified before, reducing the organization's exposure to threats.

These case studies highlight the benefits of dynamically retrainable firewalls, especially in environments where threats evolve frequently. This is due to their inability to adapt the action and detection strategies depending on generating new data sources, which is critical in today's cybersecurity landscape.

## 5.5. Performance Optimization: Strategies to Minimize Latency and Ensure Efficient Resource Use

Performance optimization is a critical consideration in the design of dynamically retrainable firewalls. In real-time applications, delays are deadly because they let threats get through the system. Hence, there is a need to give primacy to reducing latency to ensure high detection and response rates are achieved. It is possible to adopt lightweight models of machine learning that can be frequently retrained and applied within a short time without a negative impact on the speed of a system. Methods like pruning, quantizing, and knowledge distillation help optimize and simplify the machine learning models so that the threat detection process works more efficiently. Figure 5 showcases key performance optimization strategies in cybersecurity, such as caching frequently accessed data, load balancing traffic across multiple servers, utilizing content delivery networks (CDNs) to distribute content closer to users, and optimizing database queries for faster data retrieval.

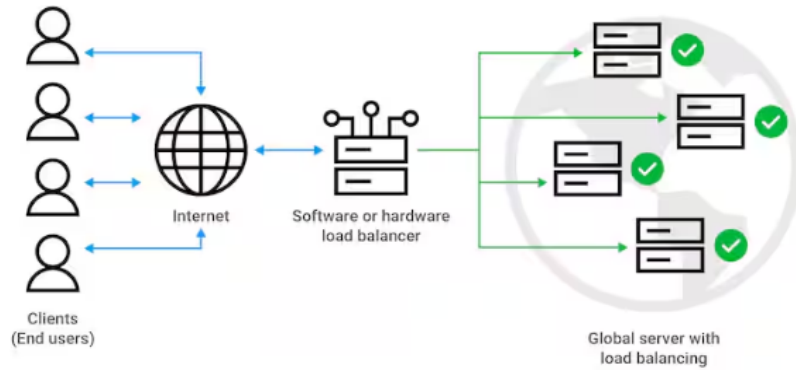

**Figure 5.** Performance Optimization.

Equation 3 calculates effective latency by accounting for processing delays $L_p$, and adaptive adjustments $A_d$, with $\alpha$ as an efficiency factor which scales the impact of adaptive delay adjustments.

$$L_{eff} = L_p - \alpha A_d \tag{3}$$

Edge computing also plays a vital role in performance optimization [36]. Through its closer proximity to the data source, edge computing means that only limited data needs to be transferred to servers for analysis. The main benefit of edge computing is minimized latency. This approach is particularly efficient in resource-sensitive environments, such as IoT networks, where bandwidth and device computational capability per device are usually low. In such circumstances, the adaptive firewalls can use edge devices to provide local threat identification of potential threats and eliminate such threats in that specific environment.

Moreover, efficient resource use is crucial to ensuring the firewall remains effective without overwhelming system capabilities. Resource allocation algorithms can help balance the computational load between the firewall's core detection and retraining processes [37]. This ensures that the firewall can dynamically allocate resources to counter new and emerging threats. Tables 3 shows a metric-based comparison between static and dynamic firewalls.

| Metric | Static Firewalls | Dynamically Retrainable Firewalls |
| --- | --- | --- |
| Latency (ms) | Low | Medium to High (optimization required) |
| Throughput (Gbps) | Moderate | High (with edge computing optimization) |
| Threat Detection Accuracy (%) | 60–75 | 85–95 |
| False Positives (%) | Higher | Lower (via ML optimization) |
| Resource Utilization (%) | Low | High |

**Table 3.** Metric-based static and dynamic firewall comparison.

*5.6. Integration with Modern Networks: Deployment in Zero Trust, Cloud, and Hybrid Environments*

The ability to integrate dynamically retrainable firewalls into modern network architectures is vital for their success. Zero Trust, cloud, and hybrid models are at the forefront of network protection, and adaptive firewalls must be able to integrate with these systems.

In a Zero Trust system, where no device or user is assumed to be trustworthy, dynamically retrainable firewalls provide an additional layer of defense by accurately identifying devices and users. These firewalls can learn the behavior patterns and flag deviations from standard behaviors as potential intrusions.

The ability of firewalls to grow and adjust to situations is paramount in cloud and hybrid environments, where networks can span multiple locations and platforms. Firewalls can be easily implemented in different cloud setups and can be updated to train newer observation patterns to look for new threats and constantly safeguard the data irrespective of its location on the cloud.

*5.7. Compatibility with Existing Systems*

Integrating dynamically retrainable firewalls with existing cybersecurity systems is another key consideration. Such firewalls should be effectively integrated with traditional security technologies, including intrusion detection and prevention systems, firewalls, and anti-virus solutions. The following compatibility helps in understanding that the new system will strengthen, as opposed to diminishing, security measures in place. Furthermore, the latest firewall decision should support the network's existing systems, which does not necessarily mean transforming the entire network system infrastructure. APIs may realize such integration and standard interfaces to complement the existing systems with the additional features of the firewall for dynamic threat identification and response.

In conclusion, while dynamically retrainable firewalls are feasible, their successful implementation depends on several key factors: architectural scalability, near real-time threat detection, the expandability of firewall throughput and the network as a whole, and the ability to integrate with modern network environments. By addressing these aspects, adaptive firewalls can provide the robust protection necessary to defend against the increasing complexity and diversity of cyber threats.

## 6. Future Trends

Emerging technologies are set to play a significant role in the future of dynamically retrainable firewalls and cybersecurity in general. One such advancement is the growth of Artificial Intelligence (AI). As machine learning algorithms continue to evolve, AI will enhance firewall capabilities in identifying and preventing real-time threats [38]. An AI-based firewall can incorporate advanced models to analyze traffic, increasing its capacity and efficiency to respond to new threats faster than conventional firewalls. Furthermore, to enhance system performance, AI will assist in perfecting the retraining process, minimizing latency, and improving resource management.

Another promising technology is quantum computing. Although currently in the prototype phase, quantum computing has the potential to revolutionize encryption and security. The underlying algorithms of quantum computers could support nearly invulnerable cryptographic systems, making it more difficult for attackers to decrypt or bypass encryption methods [39]. This could significantly enhance dynamically retrainable firewalls to substantially improve the security protocols of the next generation.

However, these advancements also present challenges. One current issue is what is right when constructing machine learning models. The integration of AI and machine learning into cybersecurity systems may entail biases in data, which in turn can lead to biased algorithms. This could result in unfair or ineffectively detected threat patterns that might not be addressed [40]. Protecting the rights of individuals involved in creating these models and guaranteeing their impartiality and clarity as the models are built will be important determinants of continued public trust in these technologies. Figure 6 depicts how AI and ML technologies are integrated into cybersecurity systems. It shows how these technologies can analyze vast amounts of data, identify patterns, and detect anomalies, enabling proactive threat prevention and rapid response to cyberattacks.

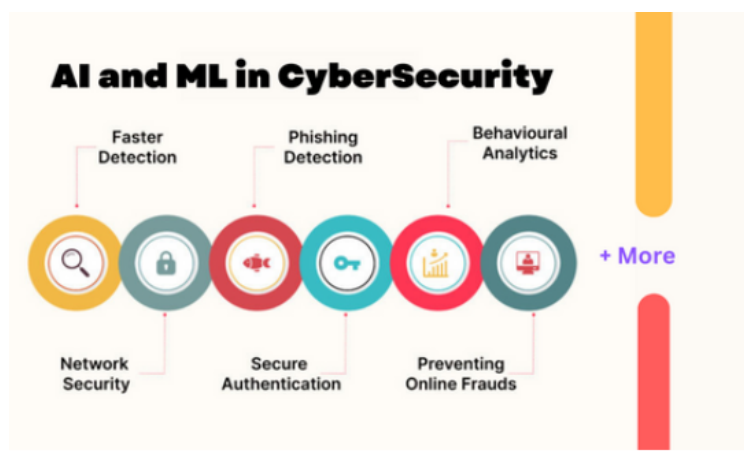

**Figure 6.** AI and ML in Cybersecurity.

Additionally, regulatory and compliance hurdles will become more complex as adaptive firewalls become widespread [41]. Governments and industry bodies will need to develop and enforce regulations to ensure AI-driven cybersecurity systems meet privacy, security, and fairness standards while allowing organizations to innovate in response to emerging threats.

## 7. Conclusion

This research has explored the evolving landscape of dynamically retrainable firewalls and their potential to revolutionize cybersecurity. The findings emphasize the significance of security infrastructures capable of reacting to advanced cyber threats. Through machine learning algorithms, firewalls that were once limited to filtering known threats can now identify previously unknown threats in real time. These system's capabilities to retrain dynamically enable it to address new types of attacks better, affording better security against new cyber threats. The disclosure of the architecture, design, and performance optimization techniques availed an obvious path for future development and improvement of these firewalls.

Another significant finding of this study is the need to design algorithms that optimize threat detection rates with system performance, especially in resource-constrained environments. Approaches such as model compression and edge computing appear feasible in resolving the issues hindering efficient functioning by offering optimization mechanisms that do not impact the threat identification performance. Further, incorporating these approaches into current network structures like the Zero

Trust model and a hybrid cloud environment highlights their potential to increase security across different organizational contexts.

The implications of this research are far-reaching. The adoption of dynamically retrainable firewalls will significantly shift how cybersecurity threats are addressed, moving from reactive to proactive measures. This shift will require changes in policy approaches for integrating adaptive technologies and the creation of new legislative frameworks to govern their practical and legal implementation. As a result, organizations can better protect their information assets, reduce threat risks, and build stronger IT environments.

In conclusion, the vision for a safer, adaptive digital infrastructure is increasingly within reach. New technologies like machine learning, artificial intelligence, and quantum computing contribute to developing better cybersecurity tools, making cyber-criminals ineffective in the organization. This research lays the groundwork for continued progress in the architecture and application of machine learning-based firewalls, offering essential insights to the cybersecurity community for creating a safer, more interactive digital society.